%% file: main.tex
\documentclass[10pt,journal,compsoc]{IEEEtran}

\ifCLASSOPTIONcompsoc
  \usepackage[nocompress]{cite}
\else
  \usepackage{cite}
\fi

\usepackage[pdftex]{graphicx}
\usepackage{booktabs}
\graphicspath{{./}}
\DeclareGraphicsExtensions{.pdf,.jpeg,.jpg,.png}
\usepackage{animate}

\usepackage{amsmath}
\usepackage[hidelinks]{hyperref}

\usepackage{xcolor}
\usepackage{listings}
\usepackage{eurosym}
\definecolor{cfgcomment}{gray}{0.45}
\lstdefinestyle{toml}{
  basicstyle=\ttfamily\footnotesize,
  columns=fullflexible,
  keepspaces=true,
  breaklines=true,
  showstringspaces=false,
  commentstyle=\color{cfgcomment}\itshape,
  morecomment=[l]{\#},
  frame=single,
  captionpos=b,
}

\usepackage{titlesec}
\titleformat{\section}{\normalfont\Large\bfseries}{\thesection}{0.5em}{}
\titlespacing{\section}{0pt}{1.5ex plus .2ex}{1.0ex plus .1ex}
\titleformat{\subsection}{\normalfont\large\bfseries}{\thesubsection}{0.5em}{}
\titlespacing{\subsection}{0pt}{1.2ex plus .2ex}{0.8ex plus .1ex}
\titleformat{\subsubsection}{\normalfont\normalsize\bfseries}{\thesubsubsection}{0.5em}{}
\titlespacing{\subsubsection}{0pt}{0.8ex plus .2ex}{0.4ex plus .1ex}

\usepackage{etoolbox}
\makeatletter
\patchcmd{\abstract}{\sffamily}{\rmfamily}{}{}
\patchcmd{\IEEEkeywords}{\sffamily}{\rmfamily}{}{}
\let\@IEEEorigmaketitle\@maketitle
\renewcommand{\@maketitle}{\begingroup\let\sffamily\rmfamily\@IEEEorigmaketitle\endgroup}
\makeatother

\begin{document}

\title{Self-replicating seedbox servers using programmable money}

\author{Matei Dogariu, Johan Pouwelse}

\IEEEtitleabstractindextext{
  \input{abstract}

}

\maketitle
\IEEEdisplaynontitleabstractindextext
\IEEEpeerreviewmaketitle

\input{introduction}
\input{proposed_solution}
\input{related_work}
\input{safety_and_security}
\input{testing_methodology}
\input{results}
\input{discussion}
\input{limitations}
\input{conclusion}

\ifCLASSOPTIONcaptionsoff
  \newpage
\fi

\input{references}
\input{appendix}

\end{document}

%% file: abstract.tex
\begin{abstract}
Centralized content distribution makes availability depend on a single operator's survival and willingness to serve. EternalSeedBox replaces the operator and network with inherited economic parameters: each node is a VPS that seeds media over BitTorrent, holds a Bitcoin wallet, and autonomously decides every twelve hours whether to renew its lease, spawn a child, or sweep its funds to a healthier peer before expiring. A single genesis node seeds the fleet, and every node thereafter is provisioned, funded, and retired autonomously. We validate the design against faithful replicas of both the Bitcoin payment network and the SporeStack VPS marketplace by running the unmodified node code. A lump sum of EUR 10,000 grew the fleet to 33 nodes before capital exhausted at day 153. With simulated income, the fleet held 40--80 live nodes across 510 days, recording 268 births and 190 deaths. A heritable caution trait was introduced to diverge across generations: low-caution lineages reproduced faster during high-income phases, while the survival advantage expected of high-caution lineages during income pauses did not appear, leaving selection in favor of low-caution nodes. The fleet tolerates high node turnover because reproduction depends on any node holding a surplus, not any single node surviving. EternalSeedBox shows that a content distribution network can lease, pay for, and replenish its own hardware without a human operator after genesis, provided income exceeds per-node rent.
\end{abstract}

%% file: introduction.tex
\IEEEraisesectionheading{\section{Introduction}\label{sec:introduction}}

\IEEEPARstart{M}{ainstream} content distribution platforms are owned and operated by single entities that control both the hardware and the material being hosted. This concentration creates single points of failure, requires continuous human supervision to remain operational, and gives the operator sole authority over enforcing copyright claims. This thesis proves the opposite is feasible: a content distribution network that runs without a central operator, which decides on its own when to grow and when to shrink. We propose EternalSeedBox, a peer-to-peer fleet of autonomous content distribution nodes that lease their own servers and earn revenue from user participation. Any node that accumulates enough funds provisions new nodes, while a node that cannot afford to sustain itself expires gracefully. Control is distributed equally across the fleet rather than held by any one party or leader node. When a node fails, the others carry on, and a new one is eventually provisioned in its place. The network continues to operate for as long as users find its services worth paying for. We propose an open market for hardware procurement, end-to-end open source software, and artists owning their creations through permissive Creative Commons licensing.

\subsection{Problem Statement}\label{sec:problem}

Centralized content distribution concentrates operational control on a small number of decision-making entities. Any such entity can remove or restrict content under commercial, legal, or political pressure. Every centralized system has this property. Whenever a single operator controls access, availability depends entirely on that operator's willingness and ability to serve it. When the operator's hardware fails, its policies change, regulators intervene, or it goes bankrupt, users lose access to the hosted material.

On such platforms, what stays online is decided by whoever runs the service. The platform keeps content available for as long as it serves its own commercial and legal interests, regardless of whether users still want it. When those interests change, content often disappears.

Decentralized alternatives address part of this problem. Peer-to-peer protocols such as BitTorrent \cite{bittorrent} and IPFS \cite{ipfs} spread content across many nodes, so content takedown is difficult. The hardware itself, however, is still owned and managed by humans, and the network only survives as long as enough keep their nodes running. Private torrent communities are often raised as a successful counterexample \cite{privatetrackers}, but they function only because human moderators enforce their membership and seeding/leeching ratio rules. What remains unsolved is whether the infrastructure layer can run without the human dependency, with the network leasing and paying for its own hardware autonomously.

EternalSeedBox is a peer-funded, autonomous content distribution network in which no single node is authoritative and no human operator manages the fleet after genesis. Each node makes its own decisions locally, based on its financial state and inherited parameters, without operator intervention. The network's continued operation depends only on users being willing to participate in its economy, with no specific node having to survive.

\subsection{Research Question}\label{sec:rq}

Can a network of autonomous software agents, seeded by a single human-initiated genesis node and controlled only by inherited economic parameters, survive and grow without further human intervention?

The question spans distributed systems, autonomous AI agents, multi-agent systems, and experimental microeconomics. We explore the engineering challenges of building a self-managed, decentralized content delivery network and proving its resilience. Frontier science within self-organizing systems manages to simulate morphology and evolutionary processes. We advance the field by running the unmodified node code against a faithful replica of both the Bitcoin network and the VPS marketplace, thus validating the system's correctness and economic dynamics before full-scale live deployment. Our scientific contribution is the first self-replicating autonomous economic agent that acquires physical infrastructure through a commercial market, operating on real payment infrastructure.

%% file: proposed_solution.tex
\section{Proposed Solution}\label{sec:proposed_solution}

EternalSeedBox is an autonomous, self-replicating content distribution network in which each node is a VPS that manages its own finances, seeds Creative Commons content over BitTorrent, and spawns child nodes when sufficiently funded. Once a genesis node is started, all subsequent nodes are provisioned, funded, deployed, and eventually expire without human involvement.

On each 12-hour tick, a node acts or does not. The choice is driven entirely by the node's own financial and operational state.

The proposed system is split into two components. The first is the fleet of EternalSeedBox nodes that produce the service, described in Section~\ref{sec:mycelium_nodes}. The second is the client application that community members run to observe the fleet and the content it serves, described in Section~\ref{sec:client}.

\subsection{EternalSeedBox Node Internals}\label{sec:mycelium_nodes}

This subsection describes a single EternalSeedBox node: how its content is sourced, how the genesis node is bootstrapped, how a running node is structured, and how nodes coordinate as a fleet.

\subsubsection{System Overview}\label{sec:overview}

Each EternalSeedBox node is a standard Linux VPS provisioned through SporeStack~\cite{sporestack}, a Bitcoin-native VPS marketplace that allows servers to be purchased, renewed, and managed entirely over the Bitcoin network and the SporeStack API. A node holds a Bitcoin wallet, seeds Creative Commons content over libtorrent~\cite{libtorrent}, participates in an IPv8 overlay network~\cite{ipv8} to exchange health telemetry with peers, and periodically evaluates whether it should top up its own runway, spawn a child, or sweep its funds before the VPS lease expires.

The only human action required is the creation of the first node, referred to here as the genesis node. After that, the fleet is self-governing. Nodes that accumulate enough financial headroom renew their VPS lease and spawn children. Nodes that run out of funds sweep their remaining Bitcoin to a peer and let the lease expire. Whether the fleet grows or shrinks depends on whether service revenue covers the running cost.

\begin{figure}[h]
    \centering
    \includegraphics[width=\linewidth]{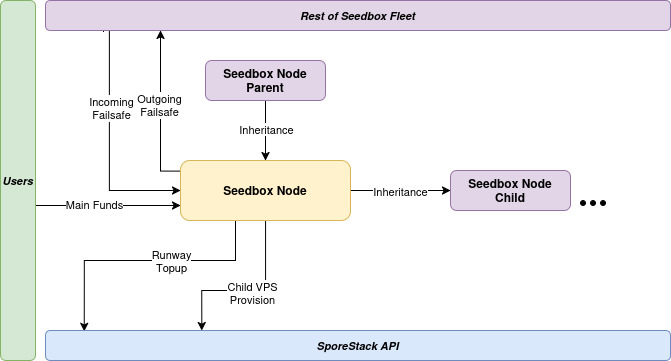}
    \caption{EternalSeedBox mode of operation. Arrows trace Bitcoin movements.}
    \label{fig:node_architecture}
\end{figure}

Figure~\ref{fig:node_architecture} traces the Bitcoin inflows and outflows with source and destination for a single node.

\subsubsection{Sourcing the Creative Commons Content}\label{sec:catalog}

A node should only distribute media whose license grants that right, so the catalog is limited to works released under the Creative Commons license. Each node receives a list of YouTube video identifiers when deployed. It contains around three million unique Creative Commons video ids.

The list is built from YouTube-Commons~\cite{youtubecommons}, a public dataset of YouTube videos published under the CC-BY 4.0 license and distributed on Hugging Face. The dataset is stored as 439 Parquet files. As the primary purpose of the dataset is to provide transcribed text, its format is not optimal for our purpose (a single video identifier appears multiple times across rows). Using the Hugging Face Hub API, each shard is read with Polars~\cite{polars}, reading only the identifier column. Concatenating the identifiers across all shards gives 22 million rows, which deduplicate to around three million distinct identifiers.

The identifiers are sorted and written to a text file. The deployment step uploads this file to every node (Section~\ref{sec:bootstrap}), and the content-download thread reads it at startup (Section~\ref{sec:seedbox}).

\subsubsection{Bootstrapping the Genesis Node}\label{sec:bootstrap}

The genesis node is created through a four-step manual process.

First, the operator creates a local HD Bitcoin wallet~\cite{bip32}. It funds the genesis VPS by purchasing SporeStack tokens, and its address is injected into every node at deployment as the cold-wallet fallback, passed down to all descendants.

A dying node that finds no live peer sends its remaining funds here rather than leaving them on an expiring VPS. The address acts as a safety net for the current experimental deployment and will be removed once the fleet runs.

Second, using the Bitcoin wallet, the operator funds a SporeStack account with tokens. An invoice is requested for the desired token amount, and subsequently Bitcoin is sent to the SporeStack payment API which returns an authentication token.

Third, the authentication token is used to provision a VPS. A fresh SSH keypair is generated, and a server with the desired specifications is launched. The server details (IP address, SSH port, and provider metadata) are saved locally.

Fourth, the deployment script connects to the provisioned VPS over SSH, installs system dependencies, configures the firewall, downloads the node code from the public git repository, generates a fresh Bitcoin mnemonic for the node's spending wallet (which is then discarded by the parent), uploads the Creative Commons video ids, and starts the orchestrator as a background process. The deployment script injects all runtime configurations as environment variables.

Each of the steps has a dedicated script: wallet management, funding the SporeStack token, acquiring the VPS, and deploying the seedbox.

\subsubsection{Node Architecture}\label{sec:architecture}
Once deployed, a node runs a single Python process, the Orchestrator. On startup, the orchestrator initializes persistent state from an SQLite database, loads the Bitcoin wallet from the injected mnemonic, sends a birth event to the external logging endpoint, and then runs multiple concurrent asynchronous tasks. The main ones are:
\begin{enumerate}
    \item \textbf{Seedbox}: a libtorrent session that creates torrent files for all content, adds them to the session, and continuously seeds them. (Section~\ref{sec:seedbox})
    \item \textbf{IPv8 announcers}: periodically broadcasts both the catalog of torrented files and the node's operational parameters.   (Section~\ref{sec:network})
    \item \textbf{Node monitor}: periodically queries the Bitcoin wallet and the SporeStack API for its operational parameters. (Section~\ref{sec:monitor})
    \item \textbf{Decision loop}: every twelve hours, evaluates the node's financial and operational state and executes one of four possible actions. (Section~\ref{sec:decision_loop})
\end{enumerate}

\subsubsection{Content Acquisition and Seeding}\label{sec:seedbox}
Each node's primary function is to seed Creative Commons media over BitTorrent. Content is downloaded at startup using \texttt{yt-dlp}~\cite{ytdlp}, a command-line tool capable of downloading content from YouTube. The node reads a text file of video IDs, downloads each file, and stores a metadata file alongside each content file. Downloads continue until disk usage reaches a configurable threshold.

The seedbox module initializes a libtorrent session listening on ports 6881 through 6891. For each file in the content directory, the seedbox creates a torrent file and adds it to the libtorrent session. The source URL and license (from the metadata file) are stored in an in-memory registry. The registry feeds both the status loop (which logs upload bytes and peer counts) and the liberation announcer (which uses the registry to broadcast torrent metadata over the IPv8 network).

The list of available torrents is sent once per new peer client connection and periodically broadcast to all peers. This keeps clients up to date with the newest torrent list and ensures that new peers receive the full catalog when they join the overlay.

\subsubsection{Fleet Communication}\label{sec:network}

Nodes communicate over an IPv8 overlay network organized as a community, identified by a fixed community ID. The community handles two message types.

\texttt{LiberatedContentPayload} (message ID 1) carries the URL, license, magnet link, and timestamp of a seeded torrent. Client connections consume this message.

\texttt{SeedboxInfoPayload} (message ID 2, WHOAMI) carries node health telemetry: friendly name, public IP, git commit hash, uptime, disk usage, Bitcoin address, Bitcoin balance in Satoshis, VPS provider region, and remaining runway in days. On the client side, this message type is needed to display the fleet and its health and wealth status. On the seedbox node side, the failsafe pipeline (Section \ref{sec:failsafe}) requires this, and future coordination protocols can reuse the same registry.

An epidemic broadcasting protocol~\cite{demers1987} is used to ensure that both message types propagate across the fleet even when two nodes are not directly connected, without excessively flooding the network. In this way we ensure that every node holds a complete view of the fleet rather than only its direct overlay neighbors.

\subsubsection{Peer Registry}\label{sec:registry}

The peer registry maintains an in-memory table of known fleet members, with Bitcoin addresses as keys. The Bitcoin address is chosen as the identity key rather than the IPv8 peer id because a SeedboxInfoPayload (WHOAMI) message forwarded by a relay arrives with the relay's id, not the originator's. The Bitcoin address remains a unique identifier across hops. Entries expire after a TTL longer than the WHOAMI broadcast interval, so a missing heartbeat eventually clears the entry.

The peer registry is the primary input to the failsafe pipeline. Dying nodes need the registry to pick a sweep target.
\subsubsection{Node Monitor}\label{sec:monitor}

The node monitor runs in the background and maintains a current reading of two quantities, the node's Bitcoin balance and its remaining VPS runway. The monitor refreshes every five minutes and exposes its output to two consumers. First, the decision loop, which uses runway to determine which action to take on tick. Second, the Seedbox Info Announcer which must include both the Bitcoin balance and the runway in every broadcast, so the node's financial state can be assessed by peers and clients.

\subsubsection{Autonomous Decision Loop}\label{sec:decision_loop}

\begin{figure}[h]
    \centering
    \includegraphics[width=\linewidth]{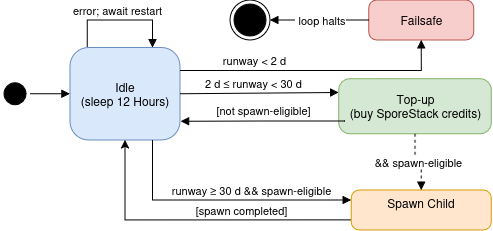}
    \caption{State machine of the decision loop.}
    \label{fig:decision_loop}
\end{figure}

The decision loop runs every twelve hours and evaluates four actions in priority order. Figure~\ref{fig:decision_loop} shows the resulting state machine.

On each tick, the node takes the first action whose guard passes. To be noted that Top-up is the one action that does not return immediately after it succeeds. Post Top-up, the node re-scans its balance and proceeds to the spawn eligibility check. A single tick can extend the lease and then spawn a child.

\textbf{Priority 1: Failsafe.} If the remaining VPS runway falls below two days, the node executes the failsafe pipeline (Section \ref{sec:failsafe}) and returns.

\textbf{Priority 2: Top-up.} If the remaining VPS runway falls below 30 days, the node computes the shortfall between its current SporeStack balance and the balance required for another 30 days of burn, then purchases enough SporeStack credit from its Bitcoin wallet to cover the lease extension. This process ensures the VPS lease remains active.

\textbf{Priority 3: Spawn.} If the node passes the spawn eligibility check (Section \ref{sec:replication}), it initiates the spawn pipeline to provision, fund, and deploy a child node.

\textbf{Priority 4: Do nothing.} If none of the above conditions triggers, the loop logs the reason and waits for the next tick.

On startup or restart, the decision loop runs a recovery procedure, shown as the self-transition on the idle state in Figure~\ref{fig:decision_loop}. It checks the persistent state database for any interrupted pipelines. If a spawn or failsafe was in progress when the process last exited, it resumes from the last logged intent of the interrupted pipeline. This guarantees that a node restart or crash does not permanently abort an irreversible ongoing operation.

\subsubsection{Self-Replication}\label{sec:replication}

The node-spawning parameters and thresholds are the key governing elements of how aggressively the fleet grows or dies. The parent must pay for the child's initial VPS lease, transfer its inheritance to the child's wallet, and account for transaction fees. A node that takes into account only its current runway could pay for the child and then bankrupt itself in the process. Our system avoids this by checking the runway the node would be left with \emph{after} the spawn, not before it. The node subtracts the full cost of the spawn from its total runway and asks whether the remainder still clears a healthy threshold. The threshold therefore represents the parent's minimum life expectancy once it has reproduced, rather than its wealth at the moment of decision. The basis for this check is the node's \emph{total} runway: pre-paid VPS days plus its SporeStack credits and Bitcoin balance that the node could still convert into VPS days. Using post-spawn total runway lets a node spawn whenever it holds the funds for a child, no matter how those funds are distributed among pre-paid VPS days, SporeStack credits, or Bitcoin balance.

The node checks two things before spawning. It must leave the node's post-spawn total runway above a threshold, and the node's Bitcoin wallet must hold enough funds to pay the child's VPS costs.

The child's inheritance is 40\% of the Bitcoin that remains once the parent has paid all the costs. A parent therefore never commits to an inheritance it cannot cover.

The threshold to spawn scales with the caution trait, a value between 0.35 and 0.9. The default value is 60 days, raised by the caution trait. A node with the default caution of 0.5 must keep 90 days of total runway after spawning. Across the caution range, the retained runway spans 81 days at 0.35 to 114 days at 0.9. The design intent is that high-caution lineages survive longer in conditions that reward conservative economic behavior, while low-caution lineages reproduce faster when there is surplus income.

When a node becomes spawn-eligible, it:

\begin{enumerate}
    \item Generates a unique child token and writes a spawn flag to the database.
    \item Mutates the caution trait. The child's trait is the parent's trait, plus a Gaussian shift with standard deviation 0.05, pulled a fraction (0.2) of the way back toward a target of 0.5.
    \item Computes the child's inheritance: 40\% of the Bitcoin balance that remains after the parent has paid the child's VPS invoice.
    \item Provisions a new VPS through SporeStack, generates a fresh Bitcoin wallet for the child, deploys the node code via SSH, injects all runtime environment variables, and starts the child's orchestrator.
    \item Transfers the inheritance amount from the parent's wallet to the child's Bitcoin address.
    \item Marks the spawn as completed and records the child's Bitcoin address in spawn history.
\end{enumerate}

The simulation component of this thesis (Section~\ref{sec:methodology}) validates these dynamics before large-scale live deployment.

\subsubsection{Caution Trait}\label{sec:caution}
Caution is the single parameter that changes from parent to child. Its purpose is to scale the post-spawn runway threshold. Lower caution lowers the threshold, so a node spawns sooner and with a smaller margin. High caution raises the threshold, so a node waits until its total runway is larger and retains more after reproducing. The genesis node starts at a value of 0.5, and every descendant mutates from its parent's value.

\subsubsection{Failsafe and Fund Safety}\label{sec:failsafe}

When the remaining runway is below two days, the node sweeps its entire Bitcoin balance to another fleet member and then waits for the VPS lease to expire naturally. A node that lets its VPS expire without sweeping loses the remaining Bitcoin permanently.

The choice of a sweep target is deliberate. The node queries its peer registry and selects the live peer with the highest current Bitcoin balance. Concentrating funds in an already-healthy peer raises the probability that the funds are not wasted.

If no live peers are visible in the registry (due to an error), the node falls back to the operator's cold wallet address injected at deployment time. A flag ensures the sweep is retried on restart if the process exits before the transaction is confirmed. After sweeping, the node continues seeding and participating in the IPv8 overlay until the lease expires. Its decision loop is halted. Without balance, none of the core decision loop actions are possible.

\subsubsection{Self-Updating Code}\label{sec:codesync}

Each node polls the remote git repository every sixty seconds. When a new commit is detected on the configured branch, the node pulls it and restarts.

The restart is safe because all durable state is stored in the database. A single commit updates the entire running fleet, with no SSH access to any node. The patch mechanism is deliberately simple, but a production deployment would need to harden it against malicious commits and broken pushes. This was useful during development, as fixing a bug or adjusting a threshold only needed a commit, with every running node picking up the change immediately. A commit-voting mechanism~\cite{stan} is being developed as a more resilient alternative.

\subsection{The Client Frontend Service}\label{sec:client}

The service produced by the fleet is distributed, so each torrent's magnet link exists only on the node that seeds it. Similarly, each node's health telemetry exists only in the broadcasts it emits. The client frontend service is a desktop application that any user can run to collect the information into one view of what the fleet seeds and its overall health.

\subsubsection{Purpose of the Client Application}\label{sec:client_purpose}

The client is a desktop application that joins the same IPv8 overlay as the fleet. It does three things for a user.

First, it presents the catalog of content the fleet seeds. The application collects the magnet link, source URL, and license of every torrent announced by the nodes, then probes the BitTorrent DHT for each one to measure swarm health.

Second, it presents fleet status. The application lists every live node with its friendly name, public IP, running git commit, uptime, disk usage, Bitcoin balance, provider region, and remaining runway. It provides a population-level picture of how many nodes are alive and their solvency.

Third, it integrates with the decentralized governance system~\cite{stan}. Community members can raise issues, propose solutions, and vote, with the resulting records distributed over IPv8. This is the voting layer that decides which code commits the fleet adopts (Section~\ref{sec:architecture}), so the client is also the interface through which the community manages the fleet.

Figure~\ref{fig:torrents_frontend_full} shows the torrent catalog view. The fleet view is shown in Appendix~\ref{app:client_img} (Figure~\ref{fig:fleet_frontend}).

\begin{figure*}[t]
    \centering
    \includegraphics[width=\textwidth]{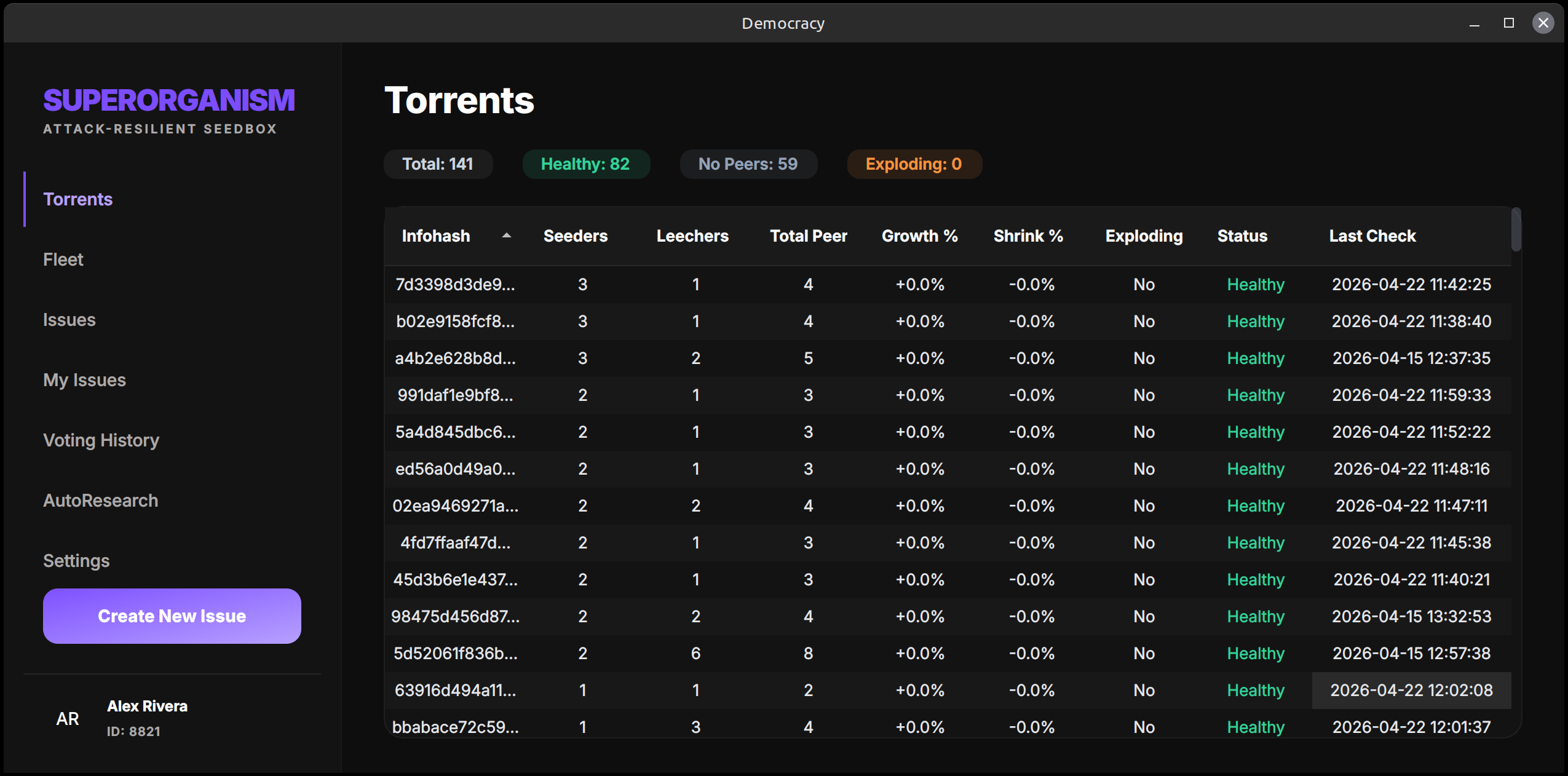}
    \caption{Torrent catalog view in the client application. Each row is a seeded torrent, with live swarm health queried from the BitTorrent DHT.}
    \label{fig:torrents_frontend_full}
\end{figure*}

\subsubsection{Node to Client Communication}\label{sec:client_data}

The client obtains its data by joining the fleet's IPv8 community. It consumes the two message types described in Section~\ref{sec:network}. \texttt{LiberatedContentPayload} (message ID 1) provides each node's catalog (URL, license, and magnet link, and infohash). The client then schedules a DHT health check for each entry. \texttt{SeedboxInfoPayload} (message ID 2) provides the fleet view: the client keeps the latest telemetry per node and filters out nodes that stop broadcasting.

The client persists all three streams (torrent catalog, fleet view, governance system) in a local SQLite database so that the view survives a restart and so that time-series metrics have historical data.

%% file: related_work.tex
\section{Related Work}\label{sec:related_work}

\subsection{Autonomic Computing}\label{sec:rw_autonomic}

Kephart and Chess argued in 2003 that growing system complexity makes human administration the primary bottleneck and that systems must self-manage to remain viable at scale~\cite{kephart2003}. Their MAPE-K reference architecture (Monitor, Analyze, Plan, Execute over a shared Knowledge base) describes a control loop in which a system observes its own state, interprets it, and adapts without operator direction. Subsequent work applied this loop to network management, storage provisioning, and adaptive middleware.

EternalSeedBox implements MAPE-K at the node level. The node monitor (Section~\ref{sec:monitor}) covers Monitor and Analyze. The decision loop (Section~\ref{sec:decision_loop}) covers Plan and Execute. The state store and inherited configuration provide the Knowledge base. The difference from prior autonomic systems is that survival depends on market conditions, as opposed to quotas set by a central operator.

\subsection{Digital Organisms and Self-Replication}\label{sec:rw_digital_organisms}

Von Neumann established that a machine carrying a complete self-description can copy that description into another and thereby reproduce~\cite{vonneumann}. One experimental system tested this in software.

Tierra~\cite{tierra} treated a segment of RAM as an evolutionary arena. Hand-coded self-replicating programs competed for memory and CPU cycles. Parasitic and symbiotic strategies were observed without external guidance. The system showed that Darwinian dynamics follow from a resource-constrained copy operation.

Replication in Tierra costs simulated CPU cycles inside a virtual environment. Replication in EternalSeedBox costs real Bitcoin transferred over a live payment network, and each new instance runs on a commercial VPS provisioned from a third-party marketplace. No prior self-replicating system has operated at this layer. The contribution of this work is the first self-replicating autonomous economic agent that acquires physical infrastructure provisioned through a commercial market.

\subsection{Decentralized Autonomous Organizations}\label{sec:rw_daos}

A Decentralized Autonomous Organization (DAO) encodes its operating rules as smart contracts on a shared distributed ledger~\cite{buterin2014}. Governance proposals are submitted as transactions, the token holders vote on-chain, with contracts executing once a quorum threshold is crossed. This removes the need for a central operator to carry out decisions.

EternalSeedBox achieves autonomous operation through a different mechanism. No shared ledger exists. Each node makes decisions from its own financial state, with no voting, no quorum, and no on-chain contract. The governance policy is the decision loop, encoded in every node. Coordination between nodes is purely economic. The absence of smart contracts removes a range of vulnerabilities and a layer of complexity, but it also means the fleet cannot reach collective decisions.

\subsection{Peer-to-Peer Economic Sustainability}\label{sec:rw_p2p_economics}

The free-rider problem is the central challenge for peer-to-peer systems that depend on voluntary resource contribution. Adar and Huberman measured Gnutella in 2000 and found that 70\% of users shared no files and the top 1\% of contributors answered nearly half of all queries~\cite{adar2000}. Cohen's BitTorrent addressed this through reciprocal choking. Peers upload preferentially to those who upload back, making sharing individually rational for anyone who also wants to download~\cite{bittorrent}.

In a live EternalSeedBox deployment, nodes earn Bitcoin from content consumers. The income incentive mechanism is left unexplored in this work, with the simulation using a mock exogenous income source. A deployment that relies on voluntary payment faces the same problem Adar and Huberman identified. The Discussion (Section~\ref{sec:Discussion}) treats this as the primary open problem for live deployment.

\subsection{Volunteer and Cooperative Content Distribution}\label{sec:rw_cdns}

Coral CDN~\cite{coral} is a cooperative content delivery network built from volunteer nodes. It routes requests to the closest cached copies, spreading load without a central operator. CoDeeN~\cite{codeen} deployed a comparable architecture on PlanetLab nodes and showed that a geographically distributed volunteer network can match commercial CDN latency across many workloads.

Both systems depend on volunteers keeping their nodes online. EternalSeedBox replaces that dependency with an economic one. A node that covers its operating cost renews its own lease, and a node that does not expires. The survival condition is changed from the operator's commitment, to market conditions. While this difference is the novel property of the EternalSeedBox design, it is also the primary risk with full-scale live deployment.

%% file: safety_and_security.tex
\section{Attack Resilience}\label{sec:attack_resilience}

A self-replicating seedbox faces a different threat model from a conventional one. Autonomous spawning, cold wallets, and peer-driven decisions create failure modes that do not exist on a single, operator-managed host. The following subsections cover the realistic attacks against each component and the residual risks the prototype accepts.

\subsection{Provisioning a Child Node}\label{sec:safety_security_provisioning}

A freshly spawned child is at the most exposed point in its lifecycle. The parent has just opened an SSH session to a fresh VPS, transferred a Bitcoin wallet, and ran the setup code on it. An attacker who intercepts any of these steps can take over a node.

The deployment script closes the child's firewall by default. The only open ports are the ones the orchestrator actually needs: SSH for parent recovery, BitTorrent for seeding, and the IPv8 overlay for fleet gossip. A scan against a freshly spawned child returns no other services. SSH uses an Ed25519 keypair generated per child, and the parent pins the host key on the first login. A man-in-the-middle attempt against a later deployment presents a different host key, fails the pinned-key check, and the parent terminates the connection before transmitting any secret. Sensitive data is piped through stdin to files written with owner-only permissions, so the mnemonic does not appear in shell history or in the output of the process status command. An attacker who later compromises a low-privilege shell on the host cannot recover credentials.

Wallet handling follows the same minimum-trace principle. The parent generates the child's wallet in a temporary database, reads out the mnemonic and address, deletes the temporary database, and only then transmits the mnemonic to the child. The parent does not keep a copy of the child's secrets, so a later compromise of the parent does not expose its children. The child reads the mnemonic from its environment once, persists it to disk with owner-only permissions, and clears the environment variable. A shell exploit on the running child cannot recover the mnemonic.

\subsection{Spawn as a Recoverable Transaction}\label{sec:safety_security_spawn_recovery}

A spawn opens four processes that cannot be undone: the SporeStack invoice, the VPS funding Bitcoin transaction, the VPS provisioning request, and the code deployment. A crash mid-spawn would put the parent node at risk of paying twice, leaving funds stranded on an expired invoice, or holding an empty VPS that does not run EternalSeedBox.

The orchestrator writes an intent record before each external call and consults it in case recovery is performed. Any Bitcoin transaction receives the strictest recovery check, since a duplicate cannot be reversed. Before re-broadcasting a funding transaction, the parent checks its own wallet history for a matching send and treats a match as proof the previous attempt succeeded. Next, SporeStack invoices are checked against an expiry buffer, so a node cannot pay an expired invoice.

After total pipeline runtime of more than two hours, spawning is abandoned. This ensures eventual node availability (even if it might incur a loss) and avoids deadlocks. The orchestrator logs the orphan identifiers, then clears the in-progress flag.

\subsection{Mitigating Runaway Behavior}\label{sec:safety_security_runtime}

Even without an external attacker, a misconfigured self-replicating system can do real damage to its own resources. Content download is set to stop once empty disk space drops below a threshold. Repeated download failures break the loop rather than retrying forever. Every outbound API call carries an explicit timeout, so a hung SporeStack call, the Bitcoin indexer, or the event logger cannot stall the main loop.

The signal handler is the other half of this. On \texttt{SIGTERM} from a lease expiry, an operator stop, or a code update, the orchestrator finishes any spawn or SporeStack top up already in flight before exiting. A node that dies in the middle of paying for a child is the worst case, so the signal handler avoids creating that case whenever possible.

The caution trait described in Section \ref{sec:replication} is inherited and mutated with each generation. Without set bounds, mutations can push nodes into one of two failure modes. A node with near-zero caution trait spawns too aggressively. A descendant with an above-one trait hoards funds and reproduces too slow. The hard clamp to $[0.35, 0.9]$ avoids both. The mutation step is Gaussian with standard deviation of 0.05, so a single generation cannot deviate significantly.

\subsection{Fleet Communication}\label{sec:safety_security_fleet}

The IPv8 overlay is the layer that a hostile or buggy peer is most likely to abuse. The orchestrator treats every inbound message as untrusted.

A peer's WHOAMI is forwarded at most once per cooldown, keyed by the originator's wallet address. A node that floods WHOAMI faster than the cooldown does not cause the rest of the fleet to amplify it. Sender metadata is rewritten on every forward hop, so the relayer's identifier in the envelope is not a trustworthy claim about who originated a message. The message body carries the originator's wallet address, which is what later forwarding decisions key on. Registry entries that have not been refreshed for an hour are dropped from queries, so the failsafe never targets a peer that has gone silent.

\subsection{Residual Risks}\label{sec:safety_security_residual}

The prototype accepts five risks rather than mitigating them.

The code update mechanism (Section \ref{sec:codesync}) does not verify commit signatures. Any push to the configured branch is pulled and executed by every live node. A compromised git account or push token therefore has the same authority as the project owner. The commit-voting mechanism~\cite{stan} is the planned replacement for this reason.

The peer registry is in-memory only, thus a restart empties it. The first failsafe tick after a restart can see no live peers and fall back to the cold wallet, even when peers are alive. Persisting the registry would mitigate this, at the cost of having to wait stale entries out on resume anyway. The prototype keeps the simpler in-memory model.

The cold wallet address used for the last-resort failsafe is injected into every node and inherited by the whole fleet. A single compromise of that address compromises this operation for the entire fleet. Finding an alternative is left to future work.

The failsafe selects the wealthiest live peer as a sweep target. This favors fleet survival, since concentrated funds are more likely to clear a spawn threshold, but it also concentrates capital in already-wealthy nodes. A successful compromise of the wealthiest visible node would attract every nearby failsafe sweep until peers updated their registries.

Inbound IPv8 messages other than WHOAMI are not rate-limited. A peer that floods \texttt{LiberatedContentPayload} is processed without restrictions. The cost is dominated by overlay bandwidth rather than persistent state, since duplicate messages do not accumulate, but a message flood would degrade a node's responsiveness.

%% file: testing_methodology.tex
\section{Methodology}\label{sec:methodology}

\subsection{Motivation of Simulation}\label{sec:motivation_simulation}

The research question of this thesis (Section \ref{sec:rq}) concerns long-term economic dynamics that a real-world deployment cannot observe within a reasonable timeline. There are three challenges that arise with real deployment.

Renting a production-ready VPS requires real funds. Ten nodes at the rate of EUR 98 per 400 GB node sum to EUR 980 per month in VPS fees, before any transfer fees. The duration of the testing is the next challenge. The Bitcoin mainnet~\cite{nakamoto} is the last, as long block times and live transaction fees slow every spawn, top-up, and failsafe operation. The research question treats all of the above as internal mechanics rather than as objects of the study.

The simulation's purpose is to remove all three constraints. It creates an environment that runs the unmodified EternalSeedBox node code without monetary cost, in compressed time, and against a reproducible and fully configurable initial state.

\subsection{Faithful Exercise of Production Code}\label{sec:faithful_execution}

The simulation aims to run the production code as much as possible. The entrypoint, the orchestrator, the wallet, and the IPv8 community are unchanged between simulation and live deployment. The changes sit at the network layer the node already calls across: the Bitcoin network and the SporeStack API\@.

A private Bitcoin network instance runs in regtest mode with a one-second block interval. Wallet creation, transaction broadcast, and balance queries go through the same Bitcoin Python library paths the production node uses, with only the network changing.

The SporeStack API is replaced by a host-side mock service. It implements the identical endpoints the production node calls: token mint, invoice quote, payment confirmation, server provision, server status, and server expiry. From the node's perspective, the mock service behaves identically to the real SporeStack API\@. Invoices are paid with real regtest Bitcoin transactions, confirmed by the local miner, and credited by the mock before the orchestrator polls again.

Each ``server'' that the mock API launches is a real LXC container on the host. Each Alpine Linux container holds its own filesystem, process tree, and IP address on the LXC bridge, and the parent deploys to that IP through the production SSH pipeline.

Peer discovery uses the same IPv8 tracker code that production nodes would use. However, a dedicated IPv8 bootstrap server is required so as to not have Internet-exposed containers. The bootstrap server is also a LXC container that boots before the genesis node.

Simulation time is scaled by a single variable. The timescale compresses remaining days, runway, and spawn thresholds. One sim-day passes in roughly 22 seconds and a 90-sim-day runway in 32 minutes. Wall-clock intervals within the nodes (heartbeat, decision loop, peer registry TTL, WHOAMI broadcasts) are also shortened when running the simulation, so concurrency, gossip, and recovery behave as on a live deployment.

Every change in the production node code only concerns simulation time constraints: shorter time intervals and more deliberate Bitcoin wallet scans due to the compressed time.

\subsection{Exogenous Income Source}\label{sec:faucet}

In live deployment, nodes earn Bitcoin from real users. The simulation models that income with a host-side income source. Per sim-day, the total payout is computed and distributed across the fleet. The payout scales with the current node count, so per-node income stays approximately invariant to population, and a uniform draw adds day-to-day variance.

The surplus payout rate is set to outweigh a limitation of the regtest server. Under the high concurrent transaction load, the regtest network drops roughly half of all broadcasts (more as simulation time grows) before confirmation, so about half of each payout never lands. The income source sends each transfer once and does not retry, because retry logic would create latency in the shared regtest server. As such, the per-day multiplier (Appendix~\ref{app:sim_config}) is set to 10 times the per-node rent to cover this loss and leave a surplus for growth. Thus, node income is well above the realistic seeding rate. The income run consequently tests fleet mechanics under favorable income.

The total payout is split across live nodes by sampling proportions from a Dirichlet distribution, producing shares that vary from node to node and from day to day. Equal shares would let no node fall behind. Uneven shares spread spawn-eligibility unevenly, with some nodes accumulating more funds than others.

The income cap (Section~\ref{sec:node_cap}) allows the fleet to see high income while the income source is active and no income during suspended intervals. One simulation run exposes the fleet to both an income phase and a drought phase. This is the basis of the caution-trait analysis in Section~\ref{sec:disc_caution}.

\subsection{Simulation Event Collection}\label{sec:event_collection}

Each node POSTs lifecycle events (births, restarts, heartbeats) to a host-side HTTP collector. The collector also records income source activity and container expiry. The event file contains all the per-run data that the analysis presented in Section \ref{sec:Results} uses.

\subsection{Host-Imposed Node Cap}\label{sec:node_cap}

Every container consumes RAM, CPU, and disk on a single host. The simulation runs on a workstation that can handle around 60 concurrent containers before unexpected errors happen. The income source enforces a sixty-node ceiling using artificial scarcity. Income is suspended when the live node count reaches 60 and resumes when it drops to 40. This is a limitation of the testing environment that could be mitigated by a more powerful machine. The evaluation in Section \ref{sec:Results} interprets results within this bound.

%% file: results.tex
\section{Results}\label{sec:Results}
\subsection{No-Income Run}\label{subsec:results_no_faucet}

In this simulation run nodes see no exogenous income, so each node spends only the runway funded at its birth and its inheritance. The genesis node was given a €10,000 lump sum at the beginning of the simulation. The run lasts 153 simulation days.

\begin{figure}[h]
    \centering
    \includegraphics[width=\linewidth]{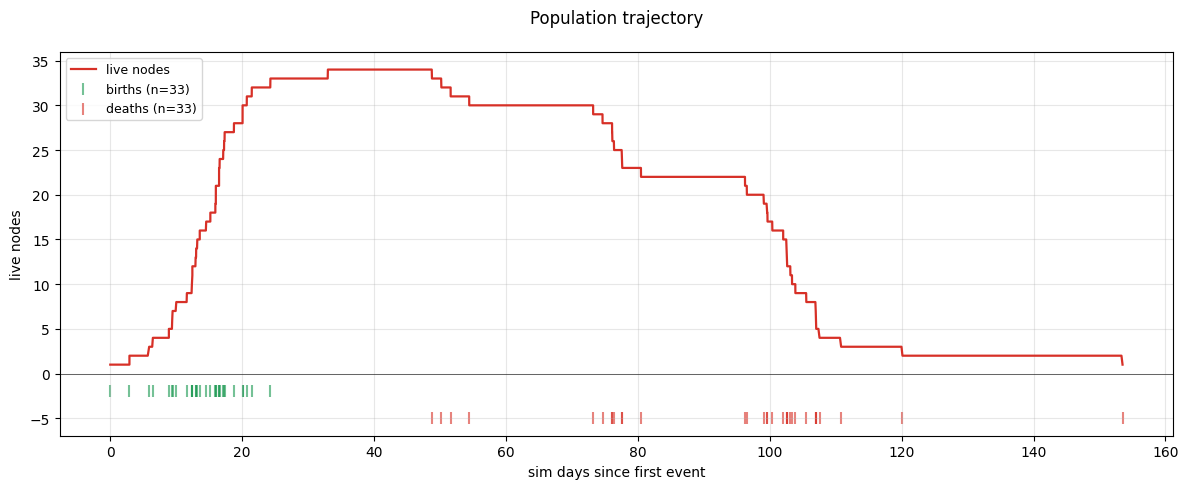}
    \caption{EternalSeedBox Population -- No-Income Run}
    \label{fig:population_no_faucet}
\end{figure}

Figure~\ref{fig:population_no_faucet} plots the live-node count over simulation time. The fleet records 33 births and 33 deaths. Every birth falls within the first 25 simulation days. The population rises to 33 live nodes and holds near that level until simulation day 48, with the first death at simulation day 49. The count then declines gradually until simulation day 95. After that day it declines steeply, and the last nodes expire at simulation day 153.

Figure~\ref{fig:lineage_no_faucet} (Appendix~\ref{app:lineage_anim}) animates every birth and death over the run, with node color representing total runway. The animation offers a visual overview of how the fleet grows and dies.

\begin{figure}[h]
    \centering
    \includegraphics[width=\linewidth]{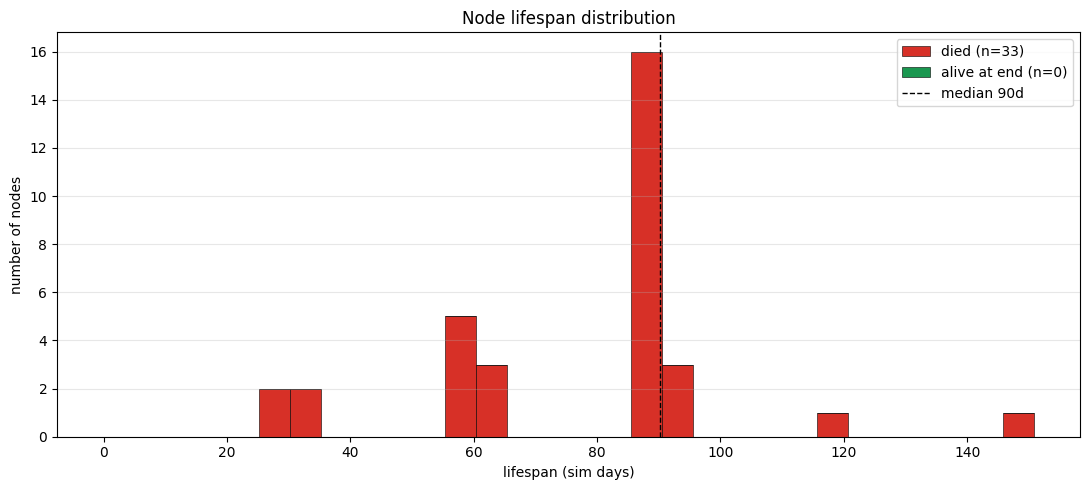}
    \caption{Lifespan Histogram -- No-Income Run}
    \label{fig:hist_no_faucet}
\end{figure}

Figure~\ref{fig:hist_no_faucet} shows the lifespan distribution across the 33 nodes. Lifespans range from 30 to 151 simulation days, with a median of 90. Sixteen nodes fall in the 88-to-92 day bin. A secondary group lives for 57 to 65 simulation days. A few nodes sit at the extremes, near 30 days at the low end and at 118 and 151 days at the high end.

\subsection{Income Run}\label{subsec:results_faucet}

In this run income is distributed to the live nodes, with fleet size artificially bounded. Income is suspended once the fleet population reaches 60 live nodes and resumed once it falls below 40. The income rate is set above any realistic seeding income to offset the regtest transaction drop (Section~\ref{sec:faucet}). Thus, this run showcases fleet mechanics under favorable, alternating income rather than economically sustainable operation. The run spans 510 simulation days.

\begin{figure}[h]
    \centering
    \includegraphics[width=\linewidth]{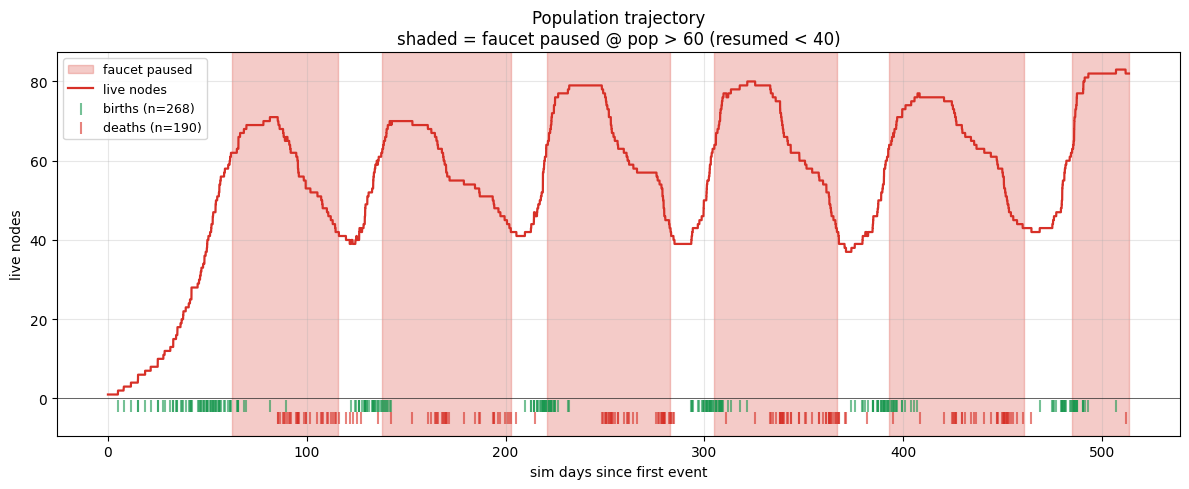}
    \caption{EternalSeedBox Population with Income Active/Suspended Markers}
    \label{fig:population_faucet}
\end{figure}

Figure~\ref{fig:population_faucet} plots the live-node count over simulation time, with the shaded bands marking the intervals where income is suspended. The run records 268 births and 190 deaths. The population climbs from the genesis node to about 70 nodes by simulation day 60, where the first pause begins. From there the count varies between 40 and 80 live nodes. Each peak of about 70 to 83 nodes coincides with a pause, and each trough of about 37 to 42 nodes coincides with a resume. The green and red marks along the baseline show the  births and deaths.

\begin{figure}[h]
    \centering
    \includegraphics[width=\linewidth]{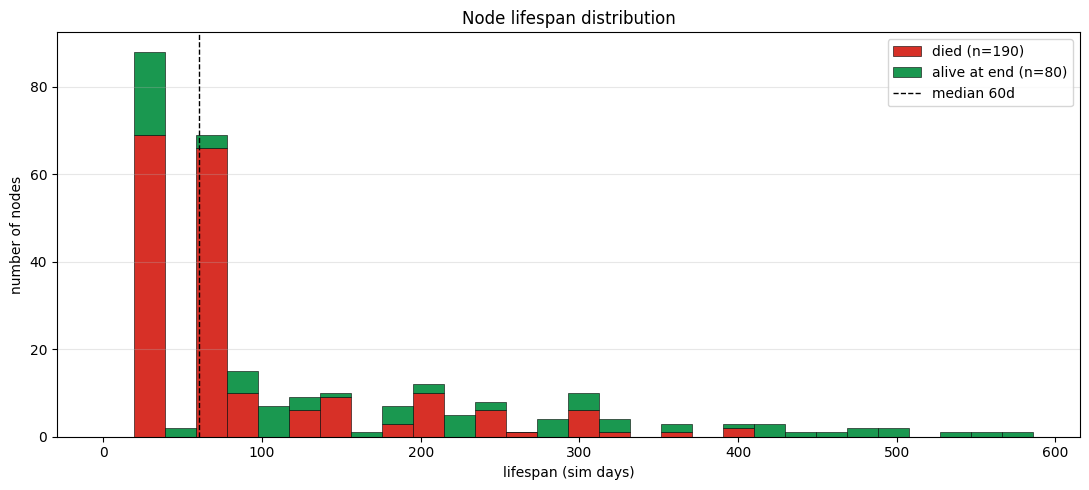}
    \caption{Lifespan Histogram}
    \label{fig:hist_faucet}
\end{figure}

Figure~\ref{fig:hist_faucet} shows the lifespan distribution across 270 nodes, split into the 190 that died and the 80 still alive at the end of the run. The median lifespan is 60 simulation days. The two leftmost bins, covering lifespans up to about 80 days, together hold roughly 160 nodes. A long tail extends past 500 days, with single nodes reaching the 500-to-600 day range. Nodes still alive at the end (green) make up most of the count in the longest-lived bins.

\begin{figure}[h]
    \centering
    \includegraphics[width=\linewidth]{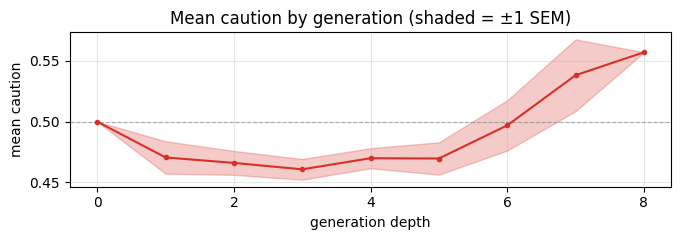}
    \caption{Caution Trait Distribution per Generation}
    \label{fig:caution_trait_distribution}
\end{figure}

Figure~\ref{fig:caution_trait_distribution} plots mean caution against generation depth, with the shaded band showing one standard error of the mean. Generation 0, the genesis node, sits at the baseline caution of 0.50. Mean caution dips to about 0.46 to 0.47 across generations 1 through 5, then rises to about 0.50 at generation 6 and to about 0.56 by generation 8. The error band widens at the younger generations, where fewer nodes contribute.

\begin{figure}[h]
    \centering
    \includegraphics[width=\linewidth]{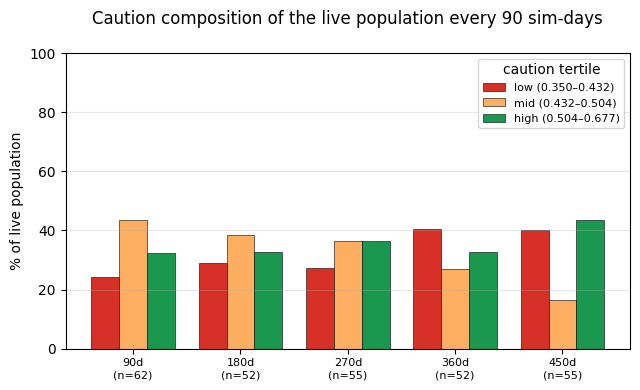}
    \caption{Caution Trait Distribution of Alive Nodes 90-day Split}
    \label{fig:caution_trait_survivorship}
\end{figure}

Figure~\ref{fig:caution_trait_survivorship} breaks the live population into caution tertiles (low 0.350 to 0.432, mid 0.432 to 0.504, high 0.504 to 0.677) at 90-day snapshot intervals. At day 90 the mid tertile holds the largest share at 43 percent. Over the following snapshots the mid share falls to about 16 percent by day 450. The low and high shares rise over the same span, and at day 450 they hold about 40 and 43 percent of the live population.

\begin{figure}[h]
    \centering
    \includegraphics[width=\linewidth]{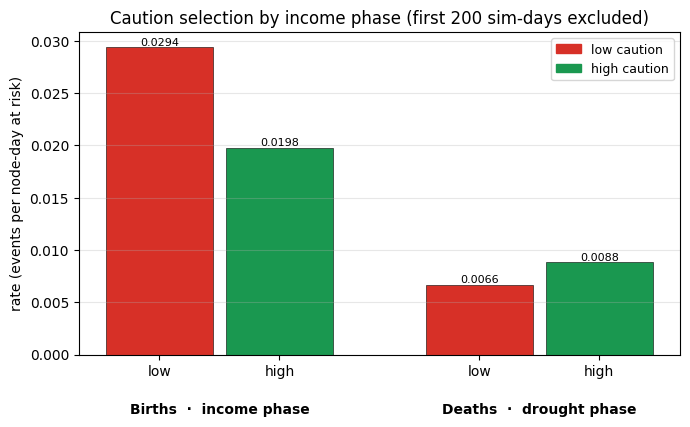}
    \caption{Birth and death rates by caution group and economic phase, first 200 simulation days excluded}
    \label{fig:caution_birth_death}
\end{figure}

Figure~\ref{fig:caution_birth_death} measures whether caution predicts reproduction and survival once the trait has diverged enough from its starting value. Each node is grouped in low or high caution by a median split of its value. Birth and death counts are converted to rates per day, computed separately for income/no-income phases. The first 200 simulation days are excluded because caution only diverges from its starting value as mutations accumulate over generations. Before the arbitrary 200-day point, the overall population caution is too close to 0.5 for an effective median split. During income phases, low-caution nodes reproduce at 0.0294 births per day against 0.0198 for high-caution nodes, a rate about 50 percent higher. During no-income phases, low-caution nodes die at 0.0066 per node-day and high-caution nodes at 0.0088.

Table~\ref{tab:storage_economics} reports the storage economics for this run, converting the simulated server spend into a real-world cost per unit of stored data.

\begin{table}[h]
\centering
\caption{Real-world storage economics for the income run}
\label{tab:storage_economics}
\begin{tabular}{lr}
\toprule
\multicolumn{2}{l}{\textbf{Real-world storage economics}} \\
\midrule
\multicolumn{2}{l}{\textbf{Parameters}} \\
1 BTC Conversion & €50,000 \\
Monthly Node Cost & €98/node/30d \\
Node Storage & 400 Gigabytes\\
\midrule
\multicolumn{2}{l}{\textbf{Sim Storage}} \\
Storage delivered & 30.7 TB-yr \\
\midrule
\multicolumn{2}{l}{\textbf{Sim Cost}} \\
Server spend (total) & €89,082 \\
Effective cost & €241 / TB-month \\
Effective cost & €2,897 / TB-year \\
Steady-state run cost & €63,271 / year \\
\midrule
\multicolumn{2}{l}{\textbf{Projected Cost}} \\
Petabyte Cost & €2,897,000 / PB-year \\
Petabyte No. Nodes & 2500 Nodes / PB-year\\
\bottomrule
\end{tabular}
\end{table}

%% file: discussion.tex
\section{Discussion}\label{sec:Discussion}

\subsection{What the Simulation Validates}\label{sec:disc_scope}

The simulation executes the unmodified seedbox node code against a faithful replica of its two external dependencies (Bitcoin payment network and the SporeStack VPS marketplace). Its purpose is to validate that the decision loop, spawn mechanics, and pipelines execute correctly. Establishing whether the revenue model holds when users replace the mock income source requires live deployment and is outside the scope of this thesis.

The two presented runs answer different questions. The no-income run isolates the spawn and inheritance mechanics from continuous income. It confirms that a funded fleet correctly expends its capital according to the spawn specification and expires on a predictable schedule. The income run tests whether the replication machinery keeps operating under sustained income and drought phases. Together they establish that the replication mechanism works as specified under both conditions.

The income source is designed to guarantee surplus. The minimum income rate to guarantee fleet survival equals the product of the node count and the per-node daily rent. The simulation sets the actual daily rate above this threshold, see Appendix~\ref{app:sim_config} Listing~\ref{lst:sim_config}. The simulation generates a high volume of Bitcoin transactions, with the single regtest node processing all of them concurrently. Under this load, some transactions are not confirmed on first broadcast. Every transfer initiated by a node is wrapped in retry logic that rebroadcasts until confirmation. On the contrary, the income source does not retry, as it would create large latency spikes. As such, the income source sends each transfer only once. To account for income distribution losses, the total income rate is set at ten times the minimum required for fleet survival. The run therefore shows that the fleet operates correctly under favorable income, not that it operates at an economically sustainable income level. Whether the system is real-world viable depends on the revenue model (Section~\ref{sec:lim_revenue}) and the break-even storage cost (Section~\ref{sec:disc_storage}), neither of which this run tries to model accurately.

\subsection{No-Income Run Lifespan Distribution}\label{sec:disc_no_faucet}

Three constants in the spawn specification determine the cluster positions in Figure~\ref{fig:hist_no_faucet}. These are the runway a node must keep after spawning, the runway each child starts with, and the inheritance ratio. Because each cluster follows from these constants, the histogram tests whether the implementation matches the specified design.

The dominant cluster at 90 days is the runway a node must keep after it spawns. A node spawns only while the runway left afterward stays above $\mathrm{spawn\_threshold} \times (1 + c)$, where $c$ is the caution trait. For the genesis value $c = 0.5$ this floor is
\[
60 \times (1 + 0.5) = 90 \text{ days.}
\]
Any node's total balance can be read in terms of runway days. One day costs EUR 3.27, which is EUR 98 spread over 30 days. Treating prepaid VPS days and wallet Bitcoin together and ignoring the small transfer fees, a parent at runway $R$ hands its child the starter runway $V = 30$ days plus 0.4 of the funds it can pass on, which leaves the parent at
\begin{align*}
    R' &= R - V - 0.4(R - 2V) \\
       &= 0.6\,R - 6 \text{ days}
\end{align*}

The parent keeps spawning while $R' \ge 90$, that is while $R \ge 160$ days. Each spawn moves the parent one step toward the floor, so a node that has spawned as many children as it can afford settles around 90 days and lives out the rest. Because nodes settle at the floor, the most common lifespan is 90 days. The observed median is 90, with 16 nodes in bin 90. Caution mutates around 0.5 across generations, so the floor $60 \times (1 + c)$ ranges from roughly 81 to 114 days, and the peak appears as a band rather than a single value.

The lower clusters at 60 and 30 days come from the last children born in each line. Each later child inherits 0.4 of a smaller balance than the one before it, so inheritance falls with every birth down a line. A child born with fewer than 160 days never spawns and lives out its starting runway $C$. The last and poorest child of a line inherits close to nothing and starts with only the 30-day starter runway $V$, while the child born just before it gets one more inheritance step and starts near 60 days. These two groups account for the 57-to-65 day cluster and the nodes near 30 days at the low end.

Two totals confirm that the fleet is a closed system. With no income, every EUR 3.27 of the inheritance funds exactly one node-day, so the fleet's total lifespan is fixed at
3061 node-days.

Spread across 33 nodes this gives a mean lifespan near 93 days, consistent with the observed median. All 33 births occur within the first 25 simulation days, before any node runs out of runway, and the remaining 128 days hold only deaths. The longest-lived node reaches 151 days because each expiring node sweeps its remaining balance to the richest live peer through the failsafe pipeline. This concentrates the remaining change in the few survivors.

Each cluster maps to a named constant. The 90-day cluster is the threshold scaled by $(1 + c)$, the 30-day cluster is the starter runway, and the 60-day cluster sits one inheritance step above it.

The no-income run therefore confirms that the spawn and inheritance logic work as intended.

\subsection{Fleet Behavior Under Sustained Income}\label{sec:disc_sustainability}

Where the no-income run exhausts a fixed capital pool in 153 days, the income run maintains a live population across 510 simulation days. While income is high, new spawns keep replacing expired nodes for as long as the run continues. Over that span, the fleet records 268 births and 190 deaths, with a live count between 40 and 80 nodes.

The lifespan distribution under income splits into two groups. Roughly 160 of the 270 nodes live fewer than 80 simulation days: their inheritance is modest and does not receive enough income to cross the top-up threshold before their runway expires. The remaining nodes form a long right tail in Figure~\ref{fig:hist_faucet}, with some reaching 500 to 600 days. These nodes received an above-median income early, and crossed the spawn threshold with a large surplus. A node that reaches the spawn threshold with a large surplus transfers 40 percent of the excess to the child and keeps the rest. Therefore, early wealth compounds. Each spawn leaves a well-funded parent able to spawn again. The long-lived nodes in the right tail all crossed the threshold carrying large surpluses.

Fleet survival does not require most nodes to be long-lived. A small fraction of well-funded nodes maintains the replication rate, continuously replacing the short-lifespan majority. The fleet accepts high turnover because reproduction depends on any nodes to hold a surplus. No individual node needs to be alive for the population to survive and reproduce.

\subsection{Population Fluctuation and the Node Cap}\label{sec:disc_fluctuation}

Figure~\ref{fig:population_faucet} shows the fleet oscillating between roughly 40 and 80 live nodes. This pattern is caused by two processes.

The income cap described in Section~\ref{sec:node_cap} suspends income when the live node count crosses 60. Likewise, income is resumed once the population falls below 40. Each pause removes all node income, allowing the low-funded population to expire and the others to deplete their reserves until income is resumed. This cycle is due to the limited capabilities of the testing hardware, which can only support around 80 concurrent containers without errors.

A second dynamic operates within the fleet itself. Each spawn moves capital from parent to child, spreading funds across a growing population. As the node count rises, the average per-node surplus above the spawn threshold shrinks, and each node becomes less likely to pass the spawn eligibility check. At the fleet sizes reached in these runs, the node cap effect is large enough to hide the internal dynamic. Testing for an organic growth cycle requires hardware capable of running several hundred concurrent containers, which would allow the cap to be removed without overwhelming the host.

\subsection{Caution Trait Evolution}\label{sec:disc_caution}

Figure~\ref{fig:caution_trait_survivorship} shows the mid-caution tertile falling from 43 percent of the live population at day 90 to 16 percent at day 450, while both the low and high tertiles grow. The mean-reverting mutation was introduced to prevent sustained drift away from the target of 0.5. Constant polarization towards the extremes requires selection pressure stronger than the intentional pull towards the center.

One explanation attributes this pressure to the income on/off cycle. Our hypothesis is that the two alternating phases reward opposite strategies. Low-caution nodes spawn aggressively and take advantage of growth phases, while high-caution nodes delay spawning and preserve runway through income-off phases. Mid-caution nodes can exploit neither phase and lose reproductive share. For this hypothesis to hold, low-caution nodes should reproduce faster during income phases, and high-caution nodes should die less often during income-off phases.

Figure~\ref{fig:caution_birth_death} tests both predictions. The reproduction prediction holds: low-caution nodes reproduce about 50 percent more than high-caution nodes while income flows (0.0294 against 0.0198 births per day). However, high-caution nodes show no survival advantage, with their death rate during no income phases being slightly higher than that of low-caution nodes (0.0088 against 0.0066 per day). Selection across the run favored low-caution nodes, because the reproduction gap operated in every income phase, while the expected high-caution survival advantage never appeared throughout the run.

The survival advantage of the high-caution nodes likely remained absent because the simulated environment was income-dominated. Income pauses are short relative to the income phases, and income resumes too early for aggressive spawning to be disadvantaged. A node that drains its runway during a brief income pause has its reserves quickly replenished once income resumes. Thus, higher caution does not show an advantage. Further testing of the hypothesis requires an environment which supports long-term scarcity and no maximum node cap. Under those conditions the high-caution survival advantage would either emerge and balance the reproduction gap, or not appear and demonstrate that caution is selectively indifferent on more cautious nodes.

\subsection{Storage Cost in Context}\label{sec:disc_storage}

Table~\ref{tab:storage_economics} reports an effective cost of EUR 241 per TB-month in the income run, projecting EUR 2.9 million per petabyte-year. Commercial object storage runs at roughly EUR 20--25 per TB-month for raw storage at market rates~\cite{cloud_storage_pricing}, a figure that covers storage capacity alone and excludes compute, bandwidth, and management fees. EternalSeedBox cannot be a competitive general-purpose storage layer. Each node rents retail VPS capacity at a fixed per-server price, unlike commercial providers that spread hardware costs across large customer bases.

EternalSeedBox's real value is availability that survives takedown demands and open-source accountability. The cost shown in Table~\ref{tab:storage_economics} therefore defines a specific market. That is, users whose content is at risk and willing to pay the premium for the availability guarantee. At EUR 98 per node per 30 days, a 400 GB node needs EUR 3.27 per day in seeding income to break even. A user storing one terabyte across 2.5 nodes would pay EUR 245 per month, matching the effective cost in Table~\ref{tab:storage_economics}. That is the minimum the fleet must earn (per node) from content consumers at current rates. Whether that pricing is acceptable for users depends entirely on the cost and incentive strategy left for future work.

%% file: limitations.tex
\section{Limitations}\label{sec:Limitations}

\subsection{Simulation Compute Constraints}\label{sec:lim_simulation}

The simulation runs each node inside an LXC container on a single host workstation. The host supports roughly 60 to 80 concurrent containers before the lack of resources causes failures. The income cap enforces a ceiling at 60 live nodes to stay within this bound. Runs on a host with capacity for several hundred containers would remove the cap and allow the population dynamics to emerge without an external pause signal. This would allow for a more natural replication cycle that is measurable as a function of the spawn and inheritance parameters.

The regtest Bitcoin network drops approximately 50 percent of broadcast transactions before confirmation. Outgoing node transactions are wrapped in extensive retry logic that guarantees completion. In contrast, the mock income source sends each transfer only once to avoid adding latency to the Bitcoin regtest server. Therefore, the surplus multiplier must compensate for the high failure rate. As such, results from the income run exaggerate the economy of a live deployment.

Simulation run duration is also limited by the memory overhead of the regtest chain scanner. The regtest miner produces one block per transaction. The node wallet scans all blocks on each balance query to derive the current UTXO set. As the chain height grows, the scanner's memory grows proportionally. As the memory requirement scales with transaction count (which itself scales with node count and simulation time), the simulation can run for a finite time until the chain cannot be held in memory anymore. As such, the income run ended once the host machine crashed.

\subsection{SporeStack as Sole VPS Provider}\label{sec:lim_sporestack}

All VPS operations call the SporeStack API. An increase in price would delay spawning across the entire fleet. An API change or service outage has the same effect but without warning. The fleet has no mechanism to redirect provisioning traffic to a secondary provider.

The simulation mock implements the same endpoints the production node calls, so a second real provider with an identical interface would require no changes to the node code. Specifying another VPS interface formally and registering a fallback provider would mitigate this external vulnerability.

\subsection{The Revenue Model}\label{sec:lim_revenue}

EternalSeedBox's operating condition is that seeding income covers per-node VPS rent. In a live deployment, this income must come from content consumers. Cohen's reciprocal choking~\cite{bittorrent} rewards upload contribution within an active transfer. Once a download completes, the incentive structure gives the client no reason to pay the seeder.

The users with incentive to pay are those who value the availability of specific content. Content unavailable on mainstream platforms has no free alternative. A user who requires such content must fund the infrastructure that hosts it. Designing a payment incentive system that exploits this demand is outside the scope of this thesis. The self-replicating node, spawn mechanics, and economic decision loop constructed in this thesis are the infrastructure layer that any later revenue mechanism can run on.

\subsection{Centralized Content Selection}\label{sec:lim_content}

Every EternalSeedBox node seeds randomly from the content catalog injected at genesis. Each spawned child inherits that same catalog. The fleet distributes content without a central coordinator, but which content it distributes is fixed by the genesis operator at launch.

The direct extension would be user content submission: a user uploads a file to the network, nodes take up seeding in response to demand signals or a vote. Two preconditions must be met before this is practical. First, nodes need an economic incentive to accept and store user-submitted files. This depends on the payment mechanism described in Section~\ref{sec:lim_revenue}. Second, a moderation protocol must handle removal requests across a network without central authority. Both are left to future work. The fixed Creative Commons catalog is appropriate for validating the economic and replication dynamics that are this thesis's primary goal.

%% file: conclusion.tex
\section{Conclusion}\label{sec:conclusion}

This thesis built EternalSeedBox, a peer-to-peer fleet of content distribution nodes that lease their own servers, pay for them in Bitcoin, and provision children once their balance is high enough. A single genesis node starts the fleet, and every node after it is deployed and retired by inherited economic parameters only. We validated the design by running the unmodified node code against a faithful replica of its two external dependencies, the Bitcoin payment network and the SporeStack VPS marketplace. The research question asked whether such a fleet can sustain and grow itself after genesis without further human involvement. When its income covers the per-node rent, it can.

The two simulation runs separate the mechanism from the economy. Without continuous income, a EUR 10,000 genesis endowment grew the fleet to 33 nodes within 25 simulation days. The fixed capital then drained on the schedule set by the spawn threshold and inheritance ratio, and the last node expired at day 153. This run confirms that nodes spend their funds exactly as per specification. With income enabled, the fleet held between 40 and 80 live nodes across 510 simulation days and recorded 268 births and 190 deaths. Most nodes were short-lived, while a small set that crossed the spawn threshold with a surplus sustained reproduction for the rest of the run. The fleet tolerates high turnover because spawning depends on any node holding a surplus rather than on any single node surviving. The inherited caution trait diverged over the run. Low-caution nodes reproduced about 50 percent faster than high-caution nodes during income phases, while the survival advantage expected of high-caution nodes during income pauses did not appear in this run. Therefore, selection favored low-caution nodes. The hypothesis that the two caution extremes complement each other remains to be further tested by a run with long income pauses and no node cap.

At EUR 50,000 per Bitcoin and EUR 98 per node per 30 days, the fleet delivered storage at EUR 241 per TB-month (Table~\ref{tab:storage_economics}), roughly ten times the rate of commercial object storage. The cause is the cost structure, with each node renting retail VPS capacity at a fixed price, whereas commercial providers lessen hardware cost across a large customer base. EternalSeedBox therefore competes on content availability and open-source transparency, not cost alone. Users who pay this premium want content that other services will not serve, thus they must fund the infrastructure that hosts it.

Two problems remain before a live deployment can replace the simulation. The first is a payment incentive (Section~\ref{sec:lim_revenue}). The live fleet needs a mechanism that makes consumers fund its services directly. The second is submission of user content (Section~\ref{sec:lim_content}). Nodes currently seed a fixed catalog of Creative Commons content, so letting users contribute content requires both the payment incentive and a moderation protocol that resolves removal requests across a fleet with no central authority. Both improvements remain to be built on top of the technical backbone that this thesis implements.

%% file: appendix.tex
\onecolumn
\appendix

\section{Lineage Animation}\label{app:lineage_anim}

\begin{figure}[ht]
    \centering
    \animategraphics[autoplay,loop,controls,poster=168,width=\linewidth]{12}{lineage_anim}{}{}
    \caption{Birth/Death events over the no-income run. Color grading represents total remaining runway.}
    \label{fig:lineage_no_faucet}
\end{figure}

\clearpage

\section{Simulation Configuration}\label{app:sim_config}

Both simulation runs reported in Section~\ref{sec:Results} share the configuration in Listing~\ref{lst:sim_config}. The two runs differ in two parameters: the no-income run (Section~\ref{subsec:results_no_faucet}) sets \texttt{faucet\_enabled = false} and \texttt{initial\_btc = 0.2}, whereas the income run (Section~\ref{subsec:results_faucet}) sets \texttt{faucet\_enabled = true} and \texttt{initial\_btc = 0.02}.

\begin{lstlisting}[style=toml,caption={Simulation configuration with income enabled (\texttt{sim\_config.toml}).},label={lst:sim_config}]
[btc]
block_interval_s = 300         # idle mining -> seconds between blocks when the mempool is empty; mempool txs are still mined on request

[sporestack]
time_scale         = 4000     # sim speed-up ratio
btc_usd            = 50000    # BTC/USD rate
monthly_cost_cents = 9800     # cents per server per 30 sim-days -> 98 eur per month for 400gb storage per node -> real SporeStack prices
invoice_lifetime_s = 600      # ttl of invoice

[genesis]
faucet_enabled          = true   # false = fleet lives on initial_btc lump sum only
days                    = 90     # SporeStack runway purchased for genesis
initial_btc             = 0.02   # regtest BTC fauceted into genesis wallet -> 1k eur ~ 0.02
faucet_max_multiplier   = 10     # daily_max = active_nodes * monthly_cost_btc * this / days_per_month
faucet_min_floor        = 0.8    # minimum = percentage of max
faucet_days_per_month   = 30     # divisor for the daily drip
faucet_pause_threshold  = 60     # pause drip when active_nodes >= this
faucet_resume_threshold = 40     # resume drip when active_nodes <= this
caution_trait           = 0.5
caution_mutation_sigma  = 0.05   # how much caution trait drift per child
caution_trait_target    = 0.5    # center the mutation drifts toward
caution_mean_reversion  = 0.2    # restoring towards mean strength; 0 = symmetric
caution_trait_min       = 0.35   # floor
caution_trait_max       = 0.9    # ceiling
spawn_threshold_days    = 60     # minimum post-spawn total runway before caution scaling (can only go up from here)
inheritance_ratio       = 0.4    # % of post-spawn remaining BTC reserves to be split with child

[intervals]
decision_interval         = 30
heartbeat_interval        = 10
peer_registry_ttl         = 120
whoami_broadcast_interval = 30
whoami_gossip_cooldown    = 30
update_check_interval     = 99999999   # disabled in sim

[network]
lxc_bridge           = "lxdbr0"
electrs_port         = 60401
event_collector_port = 8765
mock_sporestack_port = 8766
ipv8_bootstrap_port  = 7759
\end{lstlisting}

\clearpage

\section{Client Application}\label{app:client_img}

\begin{figure}[ht]
    \centering
    \includegraphics[width=\linewidth]{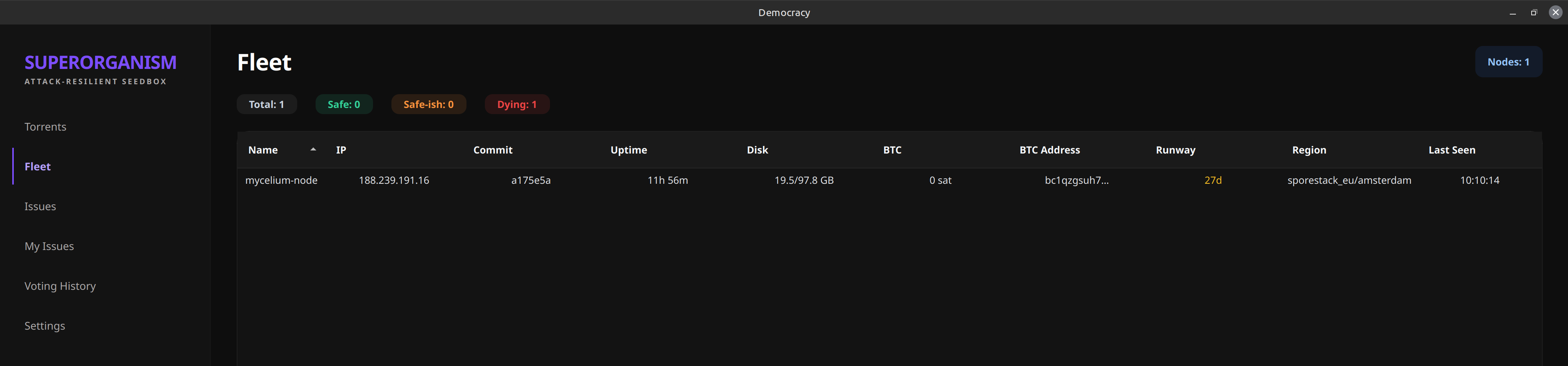}
    \caption{Fleet view in the client application. Each row is a live node.}
    \label{fig:fleet_frontend}
\end{figure}

\section{Data and Code Availability}\label{app:availability}

The EternalSeedBox node, the bootstrap tooling, and the LXC simulation harness that produced the results in Section~\ref{sec:Results} are available at \url{https://github.com/Tribler/superorganism-experiment}. The repository holds the orchestrator, the mock SporeStack service, the event collector, and the analysis scripts that generate the figures and tables reported in this work, along with the configuration in Appendix~\ref{app:sim_config}.

%% file: main.bbl
\begin{thebibliography}{99}

\bibitem{bittorrent}
B. Cohen, ``Incentives build robustness in BitTorrent,'' in
\emph{Proc. Workshop on Economics of Peer-to-Peer Systems}, Berkeley, CA,
USA, Jun. 2003. [Online]. Available:
\url{https://www.cs.princeton.edu/courses/archive/fall17/cos561/papers/BitTorrent03.pdf}.

\bibitem{ipfs}
J. Benet, ``IPFS -- content addressed, versioned, P2P file system,''
arXiv:1407.3561 [cs.NI], Jul. 2014. [Online]. Available:
\url{https://arxiv.org/abs/1407.3561}.

\bibitem{privatetrackers}
M. Meulpolder, L. D'Acunto, M. Capota, M. Wojciechowski,
J.~A. Pouwelse, D.~H.~J. Epema, and H.~J. Sips,
``Public and private BitTorrent communities: a measurement study,'' in
\emph{Proc. 9th Int. Workshop on Peer-to-Peer Systems (IPTPS~'10)},
San Jose, CA, USA, Apr. 2010. [Online]. Available:
\url{https://www.usenix.org/legacy/event/iptps/tech/full_papers/Meulpolder.pdf}.

\bibitem{vonneumann}
J. von~Neumann, \emph{Theory of Self-Reproducing Automata},
A.~W. Burks, Ed. Urbana, IL: Univ. of Illinois Press, 1966. [Online]. Available:
\url{https://cba.mit.edu/events/03.11.ASE/docs/VonNeumann.pdf}.

\bibitem{sporestack}
SporeStack, ``SporeStack: VPS hosting for Monero, Bitcoin, and Bitcoin
Cash -- no email required, API-driven,'' 2017. [Online]. Available:
\url{https://sporestack.com/}.

\bibitem{libtorrent}
A. Norberg, ``libtorrent: an efficient feature-complete C++ BitTorrent
implementation,'' 2003. [Online]. Available: \url{https://libtorrent.org/}.

\bibitem{ipv8}
Tribler, ``py-ipv8: Python implementation of Tribler's IPv8
peer-to-peer networking layer,'' 2024. [Online]. Available:
\url{https://github.com/Tribler/py-ipv8}.

\bibitem{ytdlp}
yt-dlp Contributors, ``yt-dlp: a feature-rich command-line audio/video
downloader,'' 2021. [Online]. Available: \url{https://github.com/yt-dlp/yt-dlp}.

\bibitem{youtubecommons}
PleIAs, ``YouTube-Commons: a corpus of YouTube videos shared under a
CC-BY license, with transcripts,'' Hugging Face dataset, 2024.
[Online]. Available:
\url{https://huggingface.co/datasets/PleIAs/YouTube-Commons}.

\bibitem{polars}
R. Vink \emph{et al.}, ``Polars: a fast multi-threaded, hybrid-out-of-core
DataFrame library in Rust and Python,'' 2020. [Online]. Available:
\url{https://github.com/pola-rs/polars}.

\bibitem{stan}
S. Verlaan, ``Democracy: A protocol-native decentralized incentive market
for internal system evolution,'' M.S. thesis,
Delft Univ. of Technology, Delft, The Netherlands, 2026. [Online]. Available:
\url{https://repository.tudelft.nl/}.

\bibitem{cloud_storage_pricing}
Amazon Web Services, ``Amazon S3 pricing,''
Amazon Web Services, Inc., 2026. [Online]. Available:
\url{https://aws.amazon.com/s3/pricing/}.

\bibitem{bip32}
P. Wuille, ``BIP32: hierarchical deterministic wallets,'' Bitcoin
Improvement Proposal 32, Feb. 2012. [Online]. Available:
\url{https://github.com/bitcoin/bips/blob/master/bip-0032.mediawiki}.

\bibitem{demers1987}
A. Demers, D. Greene, C. Hauser, W. Irish, J. Larson, S. Shenker,
H. Sturgis, D. Swinehart, and D. Terry, ``Epidemic algorithms for
replicated database maintenance,'' in \emph{Proc. 6th Annu. ACM Symp.
on Principles of Distributed Computing (PODC~'87)}, Vancouver, BC,
Canada, Aug. 1987, pp.~1--12. [Online]. Available:
\url{https://doi.org/10.1145/43921.43922}.

\bibitem{nakamoto}
S. Nakamoto, ``Bitcoin: a peer-to-peer electronic cash system,'' Oct. 2008.
[Online]. Available: \url{https://bitcoin.org/bitcoin.pdf}.

\bibitem{kephart2003}
J.~O. Kephart and D.~M. Chess, ``The vision of autonomic computing,''
\emph{Computer}, vol.~36, no.~1, pp.~41--50, Jan. 2003. [Online]. Available:
\url{https://doi.org/10.1109/MC.2003.1160055}.

\bibitem{tierra}
T.~S. Ray, ``An approach to the synthesis of life,'' in \emph{Artificial Life
II}, C.~G. Langton, C. Taylor, J.~D. Farmer, and S. Rasmussen, Eds.
Redwood City, CA: Addison-Wesley, 1991, pp.~371--408. [Online]. Available:
\url{https://tomray.me/pubs/alife2/Ray1991AnApproachToTheSynthesisOfLife.pdf}.

\bibitem{buterin2014}
V. Buterin, ``A next-generation smart contract and decentralized application
platform,'' Ethereum White Paper, 2014. [Online]. Available:
\url{https://ethereum.org/en/whitepaper/}.

\bibitem{adar2000}
E. Adar and B.~A. Huberman, ``Free riding on Gnutella,''
\emph{First Monday}, vol.~5, no.~10, Oct. 2000. [Online]. Available:
\url{https://doi.org/10.5210/fm.v5i10.792}.

\bibitem{coral}
M.~J. Freedman, E. Freudenthal, and D. Mazieres, ``Democratizing content
publication with Coral,'' in \emph{Proc. 1st USENIX Symp. on Networked
Systems Design and Implementation (NSDI~'04)}, San Francisco, CA, USA,
Mar. 2004. [Online]. Available:
\url{https://www.cs.princeton.edu/~mfreed/docs/coral-nsdi04.pdf}.

\bibitem{codeen}
L. Wang, V. Pai, and L. Peterson, ``The effectiveness of request redirection
on CDN robustness,'' in \emph{Proc. 5th Symp. on Operating Systems Design
and Implementation (OSDI~'02)}, Boston, MA, USA, Dec. 2002, pp.~345--360.
[Online]. Available:
\url{https://www.usenix.org/legacy/events/osdi02/tech/full_papers/wang/wang.pdf}.

\end{thebibliography}
